\documentclass[pra,superscriptaddress,twocolumn,floatfig,showpacs,floatfix]{revtex4-1}
\usepackage[boxruled]{algorithm2e}
\usepackage{graphicx} 
\usepackage{physics}
\usepackage{tensor}
\usepackage{xcolor}
\usepackage[T1]{fontenc}
\usepackage{amsmath,amssymb,bm}
\usepackage{float}
\usepackage[plainpages=false,pdfpagelabels,colorlinks=true,
linkcolor=blue,urlcolor=blue,citecolor=blue,pdftitle={Qudit-inspired optimization for graph coloring},
pdfauthor={},pdfdisplaydoctitle=true,pdfduplex=DuplexFlipLongEdge]{hyperref}
\usepackage{changes}

\date{\today}

\begin{document}

\title{
Qudit-inspired optimization for graph coloring
}

\author{David Jansen}
\affiliation{ICFO-Institut  de  Ciencies  Fotoniques,  The  Barcelona  Institute  of  Science  and  Technology, 08860 Castelldefels (Barcelona), Spain}
\author{Timothy Heightman}
\affiliation{ICFO-Institut  de  Ciencies  Fotoniques,  The  Barcelona  Institute  of  Science  and  Technology, 08860 Castelldefels (Barcelona), Spain}
\author{Luke Mortimer}
\affiliation{ICFO-Institut  de  Ciencies  Fotoniques,  The  Barcelona  Institute  of  Science  and  Technology, 08860 Castelldefels (Barcelona), Spain}
\author{Ignacio Perito}
\affiliation{ICFO-Institut  de  Ciencies  Fotoniques,  The  Barcelona  Institute  of  Science  and  Technology, 08860 Castelldefels (Barcelona), Spain}
\author{Antonio Ac\'{i}n}
\affiliation{ICFO-Institut  de  Ciencies  Fotoniques,  The  Barcelona  Institute  of  Science  and  Technology, 08860 Castelldefels (Barcelona), Spain}
\affiliation{ICREA-Institucio Catalana de Recerca i Estudis Avan\c cats, Lluis Companys 23, 08010 Barcelona, Spain}
\begin{abstract}

We introduce a quantum-inspired algorithm for graph coloring problems (GCPs) that utilizes qudits in a product state, with each qudit representing a node in the graph and parameterized by $d$-dimensional spherical coordinates. We propose and benchmark two optimization strategies: qudit gradient descent, initiating qudits in random states and employing gradient descent to minimize a cost function, and qudit local quantum annealing, which adapts the local quantum annealing method to optimize an adiabatic transition from a tractable initial function to a problem-specific cost function. Our approaches are benchmarked against established solutions for standard GCPs, showing that our methods not only rival but frequently surpass the performance of recent state-of-the-art algorithms in terms of solution quality and computational efficiency. The adaptability of our algorithm and its high-quality solutions, achieved with minimal computational resources, point to an advancement in the field of quantum-inspired optimization, with potential applications extending to a broad spectrum of optimization problems.

\end{abstract}

\maketitle

\section{Introduction}
\label{sec:intro} 
Today, many prominent tasks in industry can be formulated as integer optimization problems, such as how to efficiently use resources \cite{kallrath2005solving} or do tasks such as job scheduling \cite{ku2016mixed} or portfolio optimization \cite{he2014two}. One important example of integer optimization is the graph coloring problem (GCP): given a graph with $V$ nodes and $\mathcal{E}$ edges, assign colors to the nodes such that no two connected nodes share the same color. More generally, one might try to find the minimum number of colors needed for such an assignment to be possible, henceforth referred to as the \textit{chromatic number} of a graph ~\cite{lewis_16}. 
Being NP-complete \cite{garey1974some}, the GCP can be related to a large number of real-world problems. These include map coloring (e.g., assigning colors to provinces in Spain while avoiding that neighboring provinces have the same color), air traffic control~\cite{barnier_14}, charging electric vehicles \cite{deller_23}, resource allocation (e.g., supply chain or portfolio management), even tasks like designing seating plans, sports tournaments, or sudoku, and managing taxi bookings~\cite{lewis_16}.
Thus, there is much active research into developing algorithms that efficiently find good or optimal solutions to the GCP \textit{and} scale favorably with the size of the graph.

Quantum technologies might provide one avenue for efficiently solving optimization problems through algorithms such as the Quantum Approximate Optimization Algorithm (QAOA) \cite{farhi_14,hadfield_19,oh_19} or quantum annealing \cite{hauke_2020,pokharel_21,yarkoni_2022}. While there is a straightforward mapping from binary optimization problems to qubit systems \cite{lucas_14,glover_22}, some integer optimization problems, like GCPs, are more naturally formulated using larger local degrees of freedom. One way to accomplish this in a physics-inspired manner is to extend the notion of a qubit to a $d$-dimensional local degree of freedom, known as a \textit{qudit}. This, together with the improved capacity for simulations of physical systems with a larger number of local physical degrees of freedom, has inspired the development of hardware implementing qudit systems and qudit-inspired algorithms~\cite{lu_20,wang_20,ringbauer_22}. For instance, in Ref.~\cite{deller_23}, Deller et al. extend QAOA and integer optimization problems, including the GCP, to the qudit regime, and Ref.~\cite{bottrill_23} explores using qutrits (three-dimensional local degrees of freedom) for the GCP.

Complementary to the improvement of quantum technologies, the connection between quantum physics and optimization problems has enhanced the interest in so-called ``physics-inspired algorithms'' for optimization. A variety of such algorithms have emerged based on techniques like mean field theory \cite{smolin_14,veszeli_22,bowles_22}, tensor networks~\cite{bauer_15,mugel_22}, neural networks \cite{gomes_19,Zhao_2021,schuetz_22,khandoker_2023}, and dynamical evolution~\cite{Tiunov_19,goto_19,goto_21}. Some of these new approaches have enjoyed significant success, providing state-of-the-art solutions to existing problems as well as benchmarks to compare against possible quantum-hardware-based solutions. \\

In this work, we introduce a qudit-inspired algorithm for solving integer optimization problems. First, our approach maps the problem to a qudit system where  each qudit is parametrized by $d-$dimensional spherical coordinates. Then, we propose two schemes to solve the optimization problem.
In the first scheme, which we refer to as qudit gradient descent (QdGD), each qudit is initialized to a random state and we minimize a cost function via gradient descent.  In the second scheme, which we refer to as qudit local quantum annealing (QdLQA), we simulate the quantum annealing from the ground state of a simple Hamiltonian to that of a problem-specific cost function. This is the concept of local quantum annealing introduced by Bowles et al.~\cite{bowles_22}, in the sense that the system is constrained to remain in a product state through its evolution. We extend this idea to integer optimization, without requiring that the cost function at the end of the evolution has to have an interpretation as a quantum physical Hamiltonian. Our intuition for this approach is that we find the minima of a complicated function $f_{\textrm{final}}(\vec{\phi})$ by first initializing the parameters, $\vec{\phi}$, to a known minimum of a simple function $f_{\textrm{initial}}(\vec{\phi})$. Then, by slowly changing $f_{\textrm{initial}}( \vec{\phi} )\rightarrow f_{\textrm{final}}(\vec{\phi})$ while minimizing with respect to the parameters $\vec{\phi}$, an optimal or good solution to $f_{\textrm{final}}(\vec{\phi})$ is found. 

We test the algorithms on different standard GCPs and compare to different benchmarks in the literature~\cite{hebrard_19,schuetz_22_2,wang_23}. In almost all cases, our approach obtains results as good as or better than recently proposed state-of-the-art algorithms such as graph-neural-network-based approaches~\cite{schuetz_22_2,wang_23}. We also run our algorithms on large sparse datasets where many colors/local degrees of freedom (up to 70 \cite{hebrard_19}) are required to solve the problem. We find that even though our approaches do not find the global minimum for such high-dimensional problems, they still provide relatively good solutions (errors of the order $10^{-3}$) for problem sizes unaccounted for in state-of-the-art works such as \cite{schuetz_22_2,wang_23}. Notably, in the case of QdLQA, we find that all of these results can be obtained doing only a few gradient descent steps (for most problems only one) at each point in the adiabatic evolution, as was achieved in Ref.~\cite{bowles_22}.

The rest of this paper is structured as follows. In Sec.~\ref{sec:method}, we introduce the GCP and the algorithms. In Sec.~\ref{sec:benchmark} we benchmark our approach with other state-of-the-art methods. In Sec.~\ref{sec:conclusion}, we provide a summary and a discussion of improvements and further applications.

\section{Method}
\label{sec:method}
\subsection{Graph coloring }
\label{subsec:gc}

In our approach, we aim to solve the GCP by reframing it as finding the ground state of the antiferromagnetic Potts model~\cite{potts_1952,wu_82}. The two are closely related~\cite{wu_88,Zdeborova_07,schuetz_22_2}, and for an undirected graph $G=(V,\mathcal{E})$ with a set of nodes $V$, edges $\mathcal{E}$, and a set of $c$ colors $\{0,...,c-1\}$, the energy is given by
\begin{equation}
\label{eq:pottsmodel}
E_{\textrm{Potts}}=\sum\limits_{(i,j)\in \mathcal{E} } \delta(\sigma_i,\sigma_j) \, ,
\end{equation}
where each spin is set to one of the colors $\sigma_i \in \{0,...,c-1\}$ and $\delta(\sigma_i,\sigma_j)=1$ if $\sigma_i=\sigma_j$ and $\delta(\sigma_i,\sigma_j)=0$ otherwise. Thus, this energy is minimized whenever as many spins as possible have a different color to each of their neighbors (so that as many terms as possible are zero). The chromatic number, $\chi(G)$, is defined as the minimum number of colors $c$ required for this cost to reach zero (i.e., when no neighboring vertices share the same color). 
This problem can be posed in a number of forms, either to simply minimize the cost (NP-complete), to check whether a given graph admits a coloring with a certain number of colors (NP-complete), or to find the chromatic number of a graph (NP-hard) \cite{garey1974some}.
 
\subsection{Qudit ansatz}
\label{subsec:ansatz}

To solve GCPs, we formulate them as optimization problems for qudits. We follow Ref.~\cite{deller_23} and define the $z$ angular momentum operator for site $i$, $\hat L^z_i $, and the eigenbasis $\ket{l,m}_i$, where $l=(c-1)/2$ and $m\in \{-l,\hdots , l\}$ so that
\begin{equation}
\label{eq:Lzeigeneq}
\hat L^z_i \ket{l,m}_i=m\ket{l,m}_i \, .
\end{equation}
We also introduce
\begin{equation}
\label{eq:LP}
\hat L^+_i \ket{l,m}_i=\sqrt{(l-m)(l+m+1)}\ket{l,m+1}_i \, ,
\end{equation}
\begin{equation}
\label{eq:LM}
\hat L^-_i \ket{l,m}_i=\sqrt{(l+m)(l-m+1)}\ket{l,m-1}_i \, ,
\end{equation}
and the $x$ angular momentum operator
\begin{equation}
\label{eq:Lx}
\hat L^x_i=\frac{1}{2}(\hat L^+_i+\hat L^-_i) \, .
\end{equation}
Furthermore, we parameterize each local state by $c$-dimensional spherical coordinates using $c-1$ angles
   \begin{equation}
   \label{eq:eq1}
    \ket{\psi}_{i} = \vec{\psi}_{i}= \begin{bmatrix}
           \cos (\phi^{i}_0) \\
           \sin (\phi^{i}_0)\cos (\phi^{i}_1) \\
           \sin (\phi^{i}_0)\sin (\phi^{i}_1)\cos (\phi^{i}_2) \\
           \vdots \\
            \sin (\phi^{i}_0)\sin (\phi^{i}_1) \hdots \sin (\phi^{i}_{c-3})\cos (\phi^{i}_{c-2})\\
           \sin (\phi^{i}_0)\sin (\phi^{i}_1) \hdots \sin (\phi^{i}_{c-3})\sin (\phi^{i}_{c-2})
         \end{bmatrix} \, .
  \end{equation}
  We choose spherical coordinates for the local states since they automatically keep the state normalized and are easily differentiable, however, other parametrizations are possible (such as using Cartesian coordinates).
  
Finally, we take a product-state ansatz for the total quantum state
\begin{equation}
\label{eq:psansatz}
\ket{\psi}=\prod\limits_{i\in V} \ket{\psi}_{i} \, .
\end{equation}

\begin{figure}[t]
\includegraphics[width=0.99\columnwidth]{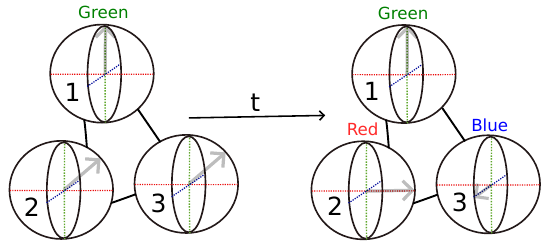}
\caption{ Three nodes (labeled 1, 2, and 3) connected by three edges. We want to assign three colors (green, red, and blue) so that no two interacting nodes have the same color. The local state is represented by a vector in a three-dimensional sphere (gray arrow). In our initial state, we assign green to node 1, and nodes 2 and 3 are in the ground states of $-\hat L^x_{2}$ and $-\hat L^x_{3}$ respectively (drawn illustratively as a superposition of the three colors). As time $t$ goes from $0$ to $1$, Eq.~\eqref{eq:timedep} is minimized with gradient descent. During the process, the spins corresponding to nodes 2 and 3 are assigned one eigenstate (color) of $\hat L^z_{2}$ and $\hat L^z_{3}$ (corresponding to the highest measurement probability) to provide a solution to Eq.~\eqref{eq:pottsmodel}. } 
\label{fig:sketch}
\end{figure}

The QdLQA approach is inspired by local quantum annealing~\cite{bowles_22}, and our goal is to start at the known minimum of one function and simulate a "quantum adiabatic evolution" into the minimum of a cost function for which the solution solves the optimization problem. During the quantum adiabatic evolution, we require that the system remains in a product state. This restriction prevents entanglement buildup in the system, which is believed to lead to the emergence of Barren plateaus~\cite{McClean_2018,Marrero_2021}. At the end of the simulation, each local state $\vec{\psi}_{i}$ is assigned a color, and Eq.~\eqref{eq:pottsmodel} is used to calculate the energy. The scheme is illustrated in Fig.~\ref{fig:sketch}.

\begin{figure}[t]
\includegraphics[width=0.99\columnwidth]{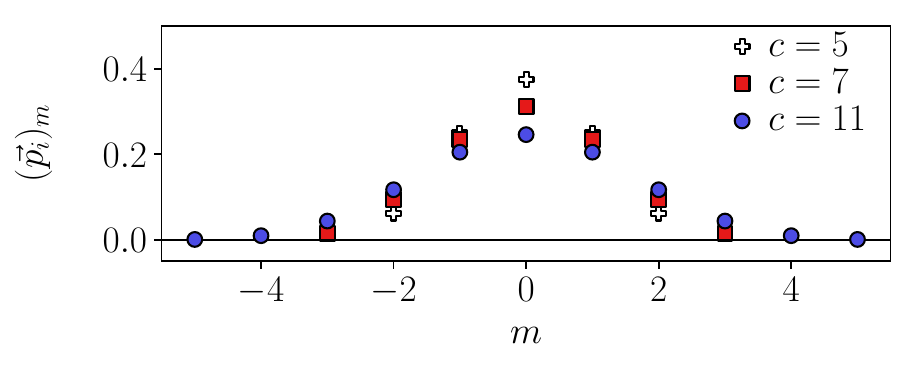}
\caption{The $m$ components of the initial distribution of the ground state of $-\hat L^x_i$ in the $\hat L^z_i$ eigenbasis. Each of the components corresponds to a different eigenvalue of  $\hat L^z_i$ [see Eq.~\eqref{eq:Lzeigeneq}], or equivalently, a different color.  We show the distribution for different numbers of colors $c$ (local Hilbert space dimension).}
\label{fig:inistate}
\end{figure}  

To simplify notation, we define the local probability vector $\vec{p}_i=\vec{\psi^2}_i$, where each component (denoted by $m$) has a clear physical interpretation, namely they give the probability of measuring the $m$-th eigenvalue of $\hat L^z_i$. This means that summing over all $m$ components for node $i$ gives $\sum\limits_m (\vec{p}_i)_m=1$. At each step in the quantum adiabatic evolution, a color configuration is generated by assigning each qudit $i$ to the state (color) corresponding to maximum probability $\text{argmax}(\vec{p}_i)$. 

We choose the initial cost function to be
 \begin{equation}
\label{eq:in}
E_{\textrm{I}}(\psi)=-\sum\limits_{i\in V  \setminus \{ j_{\textrm{max}}\}} \expval*{\hat L^x_i} \, ,
\end{equation}
where $\psi:=\psi(\vec{\phi}\,)$ is the quantum state that depends on the angles $\vec{\phi}:=(\phi^0_0,\phi^0_1,\hdots, \phi^0_{c-2}, \phi^1_0, \hdots, \phi^{\abs{V}}_{0} \hdots,\phi^{\abs{V}}_{c-2}  )$ for a system with $c$ colors and $\abs{V}$ nodes [see Eqs.~\eqref{eq:eq1} and \eqref{eq:psansatz}]. The parameters appear explicitly in the expression $\expval*{\hat L^x_i}=\tensor[_i]{\expval{ \hat L^x_i}{ \psi}}{_i}$,
and each qudit is initialized to the ground state of $-\hat L^x_i$ (which is a superposition of different colors). The initial distribution is shown in Fig.~\ref{fig:inistate}. We note that the sum in Eq.~\eqref{eq:in} goes over all qudits except $j_{\textrm{max}}$, which is the one with the most edges. This qudit is initialized with one color, e.g. $(1,0,0,\hdots, 0)^T$, and is left untouched during the adiabatic evolution. The ability to do this is due to the invariance of any solution when swapping the labels of the colors, as such, we are free to choose the color for one single spin (the one with the most edges to simplify as much as possible).

For our final cost function, we choose the scalar product between probability vectors connected by the edges. This choice arises naturally from our ansatz since two orthogonal $\vec{\psi}_i$ correspond to different colors. Using $\vec{p}_i$ instead gives the components the interpretation of a measurement probability and emits antiferromagnetic (antiparallel) solutions:
\begin{equation}
\label{eq:final}
E_{\textrm{F}}(\psi)=\sum\limits_{(i,j)\in \mathcal{E} } J_{i,j} \vec{p}_i{}^T \cdot \vec{p}_j \,,
\end{equation}
where $ J_{i,j}=1+h_{i,j}$, and $h_{i,j}$ are drawn from the uniform distribution in the interval $h_{i,j} \in [ 0,h)$ each time $E_{\textrm{F}}(\psi)$ is called. The reason for introducing the hyperparameter $h$ is to help the simulation escape local minima. If the nodes $i$ and $j$ do not have a color clash, $\vec{p}_i{}^T \cdot \vec{p}_j$ will be approximately $ 0$, and adding $h_{i,j}$ will not matter. However, for $\vec{p}_i{}^T \cdot \vec{p}_j > 0$, $h_{i,j}$ will increase the interaction strength between the nodes and encourage a change of colors. Furthermore, we note that a similar cost function (without the $h$ hyperparameter) is used the in graph neural network approaches \cite{schuetz_22_2,wang_23}. There, however, the $\vec{p}_i$ emerge from applying the softmax function to the final node embedding. Note that contrary to the algorithm of \cite{bowles_22}, this cost function does not have an interpretation as the expectation value of a physical Hamiltonian on the quantum product states considered. Alternatively, we could also have formulated a cost function corresponding to Eq.~\eqref{eq:pottsmodel} as a physical Hamiltonian in terms of polynomials in $\hat L^z_{i}$ and $\hat L^z_{j}$~\cite{deller_23}.

Furthermore, we follow Ref.~\cite{wang_23} and introduce a term to force the qudits into superpositions of all the colors,
\begin{equation}
\label{eq:weight}
E_{\textrm{W}}(\psi)=\gamma \sum\limits_{i\in V }  \vec{p}_i{}^T \cdot \log( \vec{p}_i ) \, ,
\end{equation} 
where $\gamma>0$ is a hyperparameter and the logarithm is taken component wise. For large $\gamma$, this term favors a finite probability for all possible colors, even those that are initialized close to zero in the ground state of $-\hat L^x_i $ (see Fig.~\ref{fig:inistate}). Heuristically, we find that this term helps to avoid getting stuck in local minima (similar to what was observed with graph neural networks in Ref.~\cite{wang_23}).

Finally, we conduct a quantum adiabatic evolution so that the total cost function becomes
\begin{equation}
\label{eq:timedep}
E_{\textrm{Total}}(\psi, t)=(1-t)E_{\textrm{I}}(\psi) +t\big(E_{\textrm{F}}(\psi)+E_{\textrm{W}}(\psi)\big)  \, .
\end{equation}
In the simulation, we go from $t=0 $ to $t=1$ using $N_\text{steps}$ time steps, and at each point in time, we conduct $\alpha$ gradient  descent steps. For some of the datasets discussed in Sec.~\ref{sec:benchmark}, we also find it beneficial not to start in the exact ground state of $ -L^x_i$, but rather to start in states where we perturb the initial angles in Eq.~\eqref{eq:eq1} to $\phi^i_m \rightarrow \phi^i_m +\epsilon_m$, where  $\epsilon_m$ are randomly chosen from the uniform distributed in the interval $ [ -f,f)$, and $f$ is a hyperparameter. Furthermore, we generate the classical color configuration at each time step and store the one with the lowest value for $E_{\textrm{Potts}}$ [see Eq.~\eqref{eq:pottsmodel}], which provides the best configuration found by the algorithm. 

For the alternative method focusing entirely on gradient descent, QdGD, each qudit is initialized in a random state with each qudit entry drawn from the uniform distribution in the interval $ [ 0,\tilde{f})$. The state is then normalized, and $E_{\textrm{F}}({\psi})+E_{\textrm{W}}({\psi}) $ is minimized directly by $N_\text{steps}$ gradient descent steps. We also introduce a further hyperparameter, Patience, which stops the algorithm early if no improvement in $E_{\textrm{Potts}}$ [Eq.~\eqref{eq:pottsmodel}] is seen after a certain number gradient descent steps. 

In both cases, we sample over $N_{\rm{runs}}$ runs of the algorithm and each run is stopped if $E_{\textrm{Potts}}=0$ is found. We consider the best-found solutions and their frequency as indicators of the performance of the two algorithms. 

To perform the gradient descent, we use the Adam optimizer~\cite{kingma_14} with learning $\eta$, specifically the PyTorch implementation \cite{paszke2019pytorch}. All other parameters in the Adam optimizer were set to their PyTorch default values.  Also, note that we provide pseudocode for the two algorithms in Appendix~\ref{sec:ap1}, an overview of runtimes for the calculations presented later in Appendix~\ref{sec:ap2},  and a detailed list and description of all hyperparameters specific to the algorithms in Appendix~\ref{sec:ap3}.

\section{Benchmark}
\label{sec:benchmark}

We benchmark our algorithms by considering several standard GCP instances used to test other methods~\cite{hebrard_19,li_22,schuetz_22_2,wang_23} and comparing to results in the literature~\cite{hebrard_19,schuetz_22_2,wang_23}. First, we consider Myciel and Queen graphs from the COLOR data set~\cite{colordataset_02} and the citation networks Cora and PubMed~\cite{McCallum_2000,Sen_2008,nr-aaai15}.
The Myciel and Queen graphs, based on Mycielski transformation and $n \times m$ chessboards respectively, have up to 169 nodes and densities between $16.91\%$ and $53.33\%$. In contrast, the Cora and PubMed graphs have 2708 and 19717 nodes and densities of $0.15\%$ and $0.02\%$.
Note that a more thorough description of the datasets can be found in Ref.~\cite{schuetz_22_2}.

We then consider three graphs from the SNAP library~\cite{snapnets}: The Facebook dataset based on social circles from Facebook, the Wikipedia vote dataset based on Wikipedia voting data, and the email-Eu-core data set based on email communication between members of a research institute. For these three graphs, it is known~\cite{hebrard_19} that relatively large numbers of colors are required to solve the problems (we use 70, 22, and 19) and thus they provide interesting challenges for our algorithms. In all cases, we consider undirected graphs, and we remove duplicated edges and any nonconnected nodes when prepossessing the data. 

\subsection{Influence of the hyperparameters}
\label{subsec:params}
\begin{figure}[t]
\includegraphics[width=0.99\columnwidth]{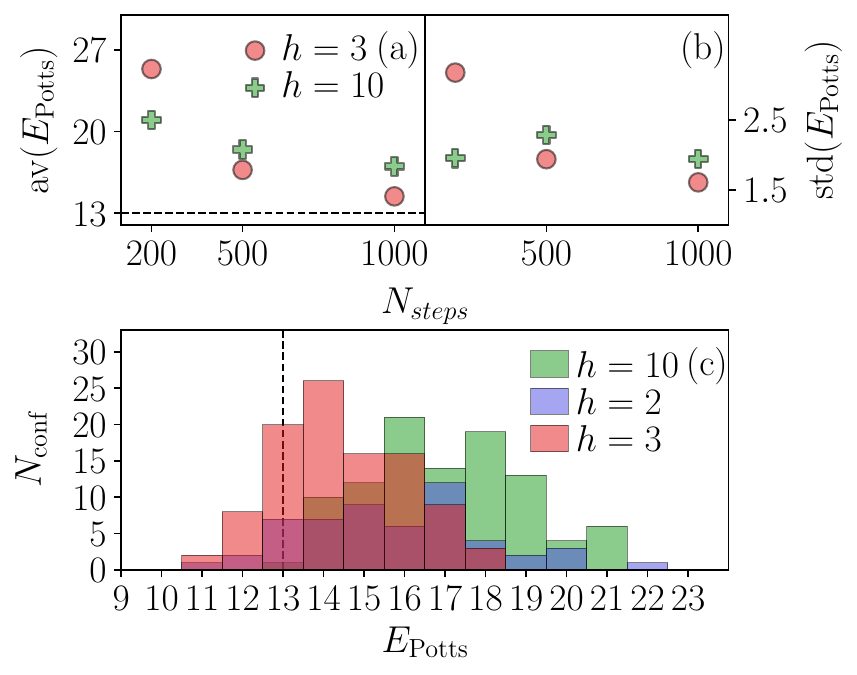}
\caption{(a) Average of color conflicts corresponding to the Potts energy from Eq.~\eqref{eq:pottsmodel} for the queen11-11 graph as a function of total number of time steps $N_\text{steps}$. We use $c=11$ colors, $f=0.0$, and different $h$. For the calculations we use $N_{\rm{runs}}=100$. The black dashed line is at 13, the best solution presented in Ref.~\cite{wang_23}. (b) Standard deviation for the data shown in (a). (c) Histogram showing the number, $N_{\textrm{conf}}$, of times a certain energy was found for $N_\text{steps}=1000$. Also here, the dashed line is at 13. All parameters not specified are provided in Appendix~\ref{sec:ap3}.}
\label{fig:averages}
\end{figure} 

\begin{figure}[t]
\includegraphics[width=0.99\columnwidth]{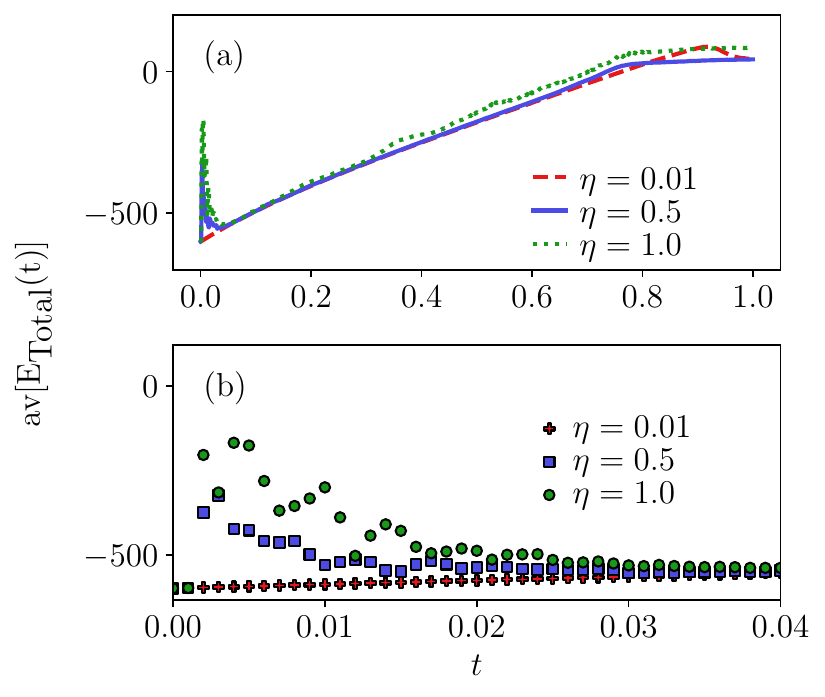}
\caption{(a) Total energy for different values of the learning rate $\eta$ for $N_\text{steps}=1000$ and $c=11$.  (b) Same data as in (a) but on a shorter time scale and showing the individual data points. All other parameters can be found in Appendix~\ref{sec:ap3}.} 
\label{fig:timeev_1}
\end{figure}

\begin{figure}[t]
\includegraphics[width=0.99\columnwidth]{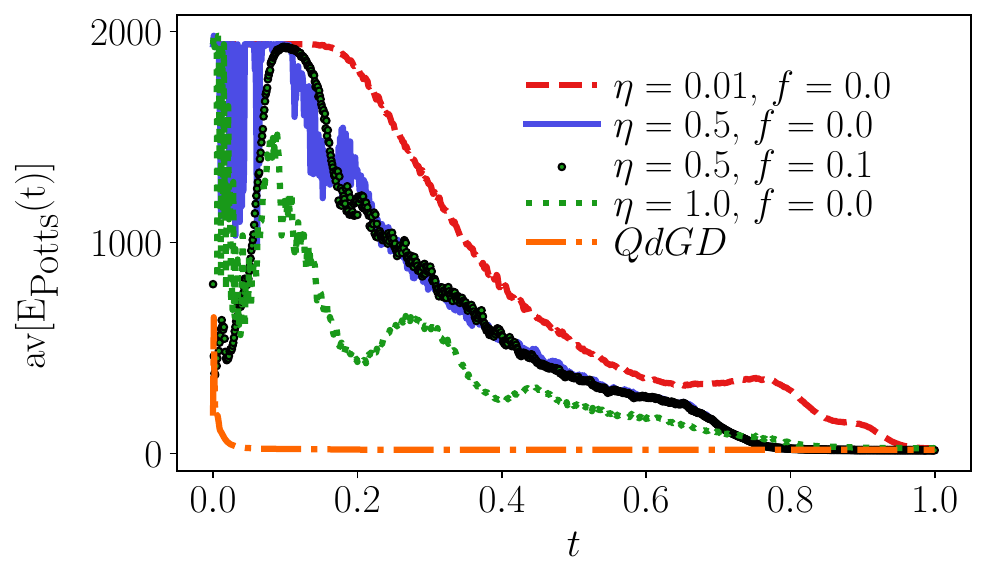}
\caption{ Average number of color clashes [$E_{\textrm{Potts}}$ from Eq.~\eqref{eq:pottsmodel}] for the same parameters as in Fig.~\ref{fig:timeev_1}. The orange line additionally shows the data calculated using QdGD and the minimum values of $E_{\textrm{Potts}}$ reached in the simulations were 19 ($\eta=0.01$, $f=0.0$), 11 ($\eta=0.5$, $f=0.0$), 10 ($\eta=0.5$, $f=0.1$), 12 ($\eta=0.01$, $f=0.0$), and 13 (QdGD) using $N_{\textrm{runs}}=100$. All other parameters can be found in Appendix~\ref{sec:ap3}. } 
\label{fig:timeev_2}
\end{figure}
First, we illustrate how some of the hyperparameters in the QdLQA algorithm influence the final solutions. As an example, we show results for the queen11-11 graph. In Fig.~\ref{fig:averages}(a), we demonstrate how the average value of the Potts energy depends on the number of time steps, $N_{\textrm{steps}}$, for different values of $h$ [which generates randomness in  $E_{\textrm{F}}({\psi})$ in Eq.~\eqref{eq:final}]. Increasing the number of steps improves the average quality of the solutions. The black dashed line shows the best value for $E_{\rm{Potts}}$ reported in Ref~\cite{wang_23}. Similar behavior is seen for the standard deviation of the $h=3$ data in Fig.~\ref{fig:averages}(b). In addition, we see that for large $N_{\textrm{steps}}$, the $h=10$ data have a larger standard deviation and average value than $h=3$. We observe that having a finite $h$ helps the algorithm escape from local minima, but if $h$ is chosen too large, this comes at the cost of more low-quality solutions. This is further illustrated in Fig.~\ref{fig:averages}(c), where we show a histogram of the number of times, $N_{\rm{conf}}$, a configuration with a certain energy is found. Again, the black dashed line shows the best value for $E_{\rm{Potts}}$ reported in Ref~\cite{wang_23}. There, we see how also choosing $h=2$ increases the fluctuations and leads to the algorithm getting stuck more frequently in a bad local minimum. In 42 instances, the solution found by $h=2$ was outside the scale of the figure. Thus, the $h=2$ data are also not plotted in Fig.~\ref{fig:averages}(a).

We now show an example of how the cost function $E_{\rm{Total}}$ and the Potts energy $E_{\rm{Potts}}$ behave during the quantum adiabatic evolution in QdLQA. In Fig.~\ref{fig:timeev_1}(a), we plot $E_{\rm{Total}}$ for the queen11-11 graph as a function of $t$ averaged over $N_{\textrm{runs}}=100$ runs for different learning rates $\eta$. There, we see that a large learning rate leads to an initial jump in the cost function, which increases when $\eta$ is increased. This can be seen better in Fig.~\ref{fig:timeev_1}(b), where the individual data points of the initial dynamics are plotted. We attribute this jump to the fact that we start in the global minimum at $t=0$, which remains close to the minimum for small increments in $t$. Thus, doing a gradient descent step with a large learning rate will make us "overshoot" and take us away from the presumptive minima. However, we see that the $\eta=0.5$ curve quickly relaxes to the $\eta=0.01$ curve for a large part of the time evolution. In contrast, the $\eta=1.0$ curve also shows large fluctuations for later times. 

Figure~\ref{fig:timeev_2} shows $E_{\rm{Potts}}$ for the same parameters as Fig.~\ref{fig:timeev_1}. Here, one can see large fluctuations in the energies for $\eta=0.5$ and $\eta=1.0$, but which decrease when $t$ is increased. The initial decay coincides with the jump in Fig.~\ref{fig:timeev_1} and can be understood as follows: In the initial state, all qudits are initialized equally, and thus, $E_{\rm{Potts}}$ has maximum value. If one now goes away from the minimum, this leads to different qudits taking on different colors, and thus the value of $E_{\rm{Potts}}$ will decrease. If one perturbs the initial state (the $f=0.1$ setting is illustrated by the black dots), the colors at different qudits will differ, and thus, $E_{\rm{Potts}}$ is smaller at $t=0$. However, at short times, the cost function will still favor all colors being the same [the ground state of Eq.~\eqref{eq:in}], and thus $E_{\rm{Potts}}$ increases. For larger times, it then closely resembles the curve for the unperturbed initial state (blue solid line) as it decreases. The orange curve in Fig.~\ref{fig:timeev_2} shows $E_{\rm{Potts}}$ for the QdGD algorithm (as a function of gradient descent steps $n=N_{\textrm{steps}} t$). Notably, this decays rapidly, which allows for a good solution to be found much earlier. When stopping the QdGD after having 100 gradient descent steps without improving $E_{\rm{Potts}}$, the average number of steps was $n=259.28 $ (corresponding to $t\approx 0.26  $).

Lastly, we want to point out another interesting feature of our approach. In most applications (e.g., often when training neural networks), optimizers like Adam update their internal parameters based on previous gradient descent steps for a fixed cost function. Here, in contrast, the parameters are updated based on the energy landscape at previous times (due to the varying cost function). 

\subsection{Comparing with other methods}
\label{subsec:litcomp}
\begin{center}
\begin{table*}
\renewcommand{\arraystretch}{1.4}
\begin{tabular}{ |c|c|c|c|c|c|c|c|c|c|c|c|c| } 
 \hline
 Graph & Nodes & Edges& Colors (c) & Tabucol &PI-CGN & PI-SAGE    &GNN-1N & $E^{QdLQA}_{\rm{Potts}}$ &$E^{QdGD}_{\rm{Potts}}$&  $p^{QdLQA}_{\textrm{min}}$ & $p^{QdGD}_{\textrm{min}} $ \\ 
 \hline
  myciel5 & 47 & 236 & 6& \textbf{0} & \textbf{0}&\textbf{0}  & \textbf{0} &\textbf{0} & \textbf{0}  & 100/100 &100/100 \\ 
 \hline
   myciel6 & 95 & 755 & 7& \textbf{0} & \textbf{0}& \textbf{0} & \textbf{0} &\textbf{0} & \textbf{0}  & 38/100  &97/100 \\ 
 \hline
 queen5-5 & 25 & 160 & 5 & \textbf{0} &  \textbf{0}&\textbf{0} & \textbf{0} &\textbf{0} &\textbf{0} & 100/100  &68/100 \\ 
 \hline
 queen6-6 & 36 & 290 & 7& \textbf{0} &1& \textbf{0} & \textbf{0} &\textbf{0} &\textbf{0} &12/100&7/100 \\ 
 \hline
 queen7-7 & 49 & 476 & 7& \textbf{0} & 8&\textbf{0} & \textbf{0} &\textbf{0} &\textbf{0} & 17/100 &8/100  \\ 
 \hline
 queen8-8 & 64 & 728 & 9& \textbf{0} & 6& 1 &1& \textbf{0} &  \textbf{0}& 6/100  &2/100 \\ 
 \hline
 queen9-9 & 81 & 1056 & 10& \textbf{0} &13& 1 & 1 &\textbf{0} &\textbf{0} & 3/100 &1/100  \\ 
 \hline
 queen8-12 & 98 & 1368 & 12& \textbf{0} &10& \textbf{0} & \textbf{0} &\textbf{0} &\textbf{0} & 27/100 &41/100 \\ 
 \hline
 queen11-11 & 121 & 1980 & 11& 20 &37& $13^{*}$ & 13 &\textbf{10} &13& 1/100 &2/100  \\ 
 \hline
 queen13-13 & 169 & 3328 & 13& 25 &61& 26 &15 &\textbf{12} &15&2/100 &1/100 \\ 
 \hline
 Cora & 2708 & 5429 & 5& \textbf{0} & 1 & \textbf{0}&X &1 &\textbf{0} &3/100  & 1/100 \\ 
 \hline
 Pubmed & 19717 &44338 & 8& NA &\textbf{13} &  17&X&15 &27 & 3/100 & 3/100 \\ 
 \hline
\end{tabular}
\caption{Datasets from \cite{colordataset_02} and the citation graphs~\cite{McCallum_2000,Sen_2008,nr-aaai15} introduced in the main text. Here, ``Colors'' represents the number of colors used in the calculations, while the following values are related to the cost function obtained after optimization by various methods, such that zero is optimum. Values given for Tabucol~\cite{schuetz_22_2}, PI-CGN \cite{schuetz_22_2}, PI-SAGE~\cite{schuetz_22_2}, and GNN-1N~\cite{wang_23} represent other algorithms, whilst the quantities related to $E^{\textrm{QdLQA}}_{\text{Potts}}$ and $E^{\textrm{QdGD}}_{\text{Potts}}$ show the best solution found by the corresponding method in $N_{\textrm{runs}}$ runs. The parameters $p^{\textrm{QdLQA}}_{\textrm{min}}$ and $p^{\textrm{QdLQA}}_{\textrm{min}}$ give the fractions, $N_{\textrm{min}}/N_{\textrm{runs}} $, indicating how often, $N_{\textrm{min}}$, the solution was found in the $N_{\textrm{runs}}$ runs. The best solution found by any of the algorithms is indicated in bold face, NA means that no solution was found within 24 h~\cite{schuetz_22_2}, and X means no value was provided in Ref.~\cite{wang_23}. For Cora and PubMed, the choice of $c$ follows Refs.~\cite{li_22,schuetz_22_2}. * 17 was reported in~\cite{schuetz_22_2} but we find 13 in our own calculations, see Appendix~\ref{sec:ap2}.
}
\label{table:t1}
\end{table*}
\end{center}
We now compare the solutions found by our algorithms to previously published results using graph neural networks in Refs.~\cite{schuetz_22_2,wang_23}. In Table~\ref{table:t1}, we show the energies of the best configurations found with QdLQA and QdGD ($E^{QdLQA}_{\rm{Potts}}$ and $E^{QdGD}_{\rm{Potts}}$, respectively) and the fraction of the runs that found that energy ($p^{QdLQA}_{\textrm{min}}$ and $p^{QdGD}_{\textrm{min}} $) for different graphs. Furthermore, we show the corresponding values reported in Refs.~\cite{schuetz_22_2,wang_23} obtained using a Tabucol~\cite{hertz_87} and a greedy algorithm, a physics-inspired graph neural network solver (PI-SAGE)~\cite{schuetz_22_2}, and graph neural networks using a first-order negative message passing strategy (GNN-1N) \cite{wang_23}. Note that if $E_{\rm{Potts}}=0$, the graph was colored with no conflicts, otherwise, $E_{\rm{Potts}}>0$ gives the remaining number of color clashes.
For all the graphs except queen11-11, queen13-13, Cora, and PubMed, both our algorithms find the optimal solution. However, some graphs seem to be particularly hard. For example, for the queen8-8 and queen9-9 graphs, both algorithms only find the optimal solutions in less than $10$ of the 100 runs. For queen11-11 and queen13-13, QdLQA finds better solutions than those reported in Refs.~\cite{schuetz_22_2,wang_23}. In the case of queen11-11, it additionally finds $E_{\rm{Potts}}=11$ in three of the 100 runs, and both $E_{\rm{Potts}}=12$ and $E_{\rm{Potts}}=13$ in nine of the 100 runs each. For queen13-13, it additionally finds $E_{\rm{Potts}}=13$ in two, $E_{\rm{Potts}}=14$ in two, and $E_{\rm{Potts}}=15$ in seven of the 100 runs.

By systematically increasing the number of colors used, we can get upper bounds for the chromatic number $\chi(G)$ which we compare to those reported in Ref.~\cite{schuetz_22_2}. For queen11-11, both our algorithms find $\chi_{\mathit{Upper}}(G)=13$ with $p^{QdLQA}_{\textrm{min}}=37/100$ and $p^{QdGD}_{\textrm{min}}=22/100$ (14 is found by the PI-SAGE algorithm in Ref.~\cite{schuetz_22_2}, 11 is optimal) and for queen13-13, both find $\chi_{\mathit{Upper}}(G)=15$ with $p^{QdLQA}_{\textrm{min}}=43/100$ and $p^{QdGD}_{\textrm{min}}=3/100$ (17 is found by the PI-SAGE algorithm in Ref.~\cite{schuetz_22_2}, 13 is optimal). 

For the Cora graph, only the QdGD algorithm finds the $E_{\rm{Potts}}=0$ solution, and only in one of the 100 runs. QdLQA finds $E_{\rm{Potts}}=1$ configurations in three and $E_{\rm{Potts}}=2$ in 20 of the 100 runs, but requires a comparatively large $h$ to do so (we use $h=10$). In this case, one must compromise between finding $E_{\rm{Potts}}=1$ a few times at the cost of obtaining many bad solutions. For example, using $h=3$ instead, we get $E_{\rm{Potts}}=2$ in 58 of 100 runs, but never $E_{\rm{Potts}}=1$. 

For the PubMed graph, neither algorithm finds better solutions than the best one reported in Ref.~\cite{schuetz_22_2}, but QdLQA seems to do significantly better than QdGD. In addtion, in contrast to the other datasets in Table~\ref{table:t1}, we do not fix the node with the largest number of edges for the PubMed graph. Rather, we fix a node with one edge as this improves the results. For $N_{\textrm{runs}}=100$, QdLQA improves from finding $E^{QdLQA}_{\textrm{Potts}}=17$ three times to finding $E^{QdLQA}_{\textrm{Potts}}=15,$ $16,$ and $17$ three times, onece, and seven times, respectively. QdGD improves from finding $E^{QdGD}_{\textrm{Potts}}=28$ twice to finding $E^{QdGD}_{\textrm{Potts}}=27$ three times. This indicates that cleverly fixing one (or more) nodes might lead to further improvements.

Note that we provide a comprehensive overview of the average runtimes, computer architectures used, and a more thorough comparison to PI-SAGE in Appendix~\ref{sec:ap2} and a list of parameters used in Appendix~\ref{sec:ap3}.
 \subsection{Benchmarking using large number of colors}
\label{subsec:largcec}
\begin{center}
\begin{table*}
\renewcommand{\arraystretch}{1.4}
\begin{tabular}{ |c|c|c|c|c|c|c|c|c|c|c|c| } 
 \hline
 Graph & Nodes & Edges& $\chi_{\mathit{Lower}}(G)$ & $\chi_{\mathit{Upper}}(G)$ & $E^{QdLQA}_{\rm{Potts}}$ & $E^{QdGD}_{\rm{Potts}}$ &   $p^{QdLQA}_{\textrm{min}}$ & $p^{QdGD}_{\textrm{min}} $ \\ 
 \hline
 email-Eu-core & 986 & 16064 & 19& 19 & 26 & 30 & 1/100& 1/100 \\ 
 \hline
 Facebook & 4039 & 88234 & 70& 70 & 25 & 22 & 1/100& 1/100\\ 
 \hline
 Wiki-vote & 7115 &103689 & 19& 22 &192 &  207 & 1/100& 1/100\\ 
 \hline
\end{tabular}
\caption{Minimum energy, $\min (E_{\rm{Potts}})$, found for the email-Eu-core, Facebook, and Wiki-vote datasets~\cite{snapnets} together with the upper and lower bounds for the chromatic number $\chi(G)$ reported in Ref.~\cite{hebrard_19}. In our simulations, we set the number of colors $c=\chi_{\mathit{Upper}}(G)$ from \cite{hebrard_19}. The parameters used can be found in Appendix~\ref{sec:ap3}.}
\label{table:t2}
\end{table*}
\end{center} 
Finally, we test our algorithms on graphs where it is known that relatively large numbers of colors are needed to solve the problems~\cite{hebrard_19}. In Table~\ref{table:t2}, we show results for the previously mentioned email-Eu-core, Facebook, and Wiki-vote datasets together with the corresponding upper and lower bounds provided in Ref.~\cite{hebrard_19} (the problem was solved for email-Eu-core and Facebook). In our simulations, we use the number of colors equal to the upper bounds provided in Ref.~\cite{hebrard_19} (19 for email-Eu-core, 70 for Facebook, and 22 for Wiki-vote), and again, $E^{QdLQA}_{\rm{Potts}}$ and $E^{QdGS}_{\rm{Potts}}$ give the number of conflicts in the best solution found. Even though neither algorithm solves the problem completely, they are clearly capable of providing relatively good solutions (quantified by the normalized error defined as the energy divided by number of edges, $\epsilon=E_{\rm{Potts}}/\abs{\mathcal{E}}$~\cite{schuetz_22_2}). For the email-Eu-core, Facebook, and Wiki-vote datasets, $\epsilon\approx 0.0016$ (0.0019), 0.00028 (0.00025) and 0.0019 (0.002) for QdLQA (QdGS). This demonstrates the ability of the algorithms to find good solutions for relatively large graphs requiring many colors on a simple laptop or desktop (see Appendix~\ref{sec:ap2} for details). 

\section{Conclusion}
\label{sec:conclusion}
In this work, we have proposed two quantum-inspired algorithms to efficiently solve integer optimization problems. In our approach,  we map nodes to qudits parameterized by $c$-dimensional spherical coordinates and perform either local quantum annealing (QdLQA) or a gradient descent (QdGD). 

We compared our results for a range of GCPs to some of the state-of-the-art methods in the literature~\cite{schuetz_22_2,wang_23}, and both QdLQA and QdGD find equally good or better solutions for all of the test cases, with the exception of the PubMed graph. 
We also tested the algorithms on graphs with large chromatic numbers~\cite{hebrard_19}. Even though neither algorithm solved these problems completely, they still provided reasonable-quality solutions (with normalized error $\epsilon\approx 10^{-3}$) using modest computing resources such as a desktop or a laptop. Notably, QdLQA almost always provided  equally good or better solutions than QdGD (with the exception of the Cora and the Facebook graph). However, QdGD was heuristically found to be more efficient, since it minimizes the desired cost function directly. We refer the reader to Appendix~\ref{sec:ap2} for further details. 

There are many compelling possible extensions to our work. For example, each local Hilbert space dimension can be set manually, and thus, the approach can easily be extended to integer optimization problems requiring different numbers of states on each node. It general, it would also be interesting to do a detailed comparison between local quantum annealing and simulated (thermal) annealing; however, it seems that for the GCP, the latter suffers convergence and scalability issues~\cite{kose2017simulated,nolte1996simulated,lukasik2008parallel}. Furthermore, the interpretation of a vector on a sphere might provide new insights into why certain solutions are preferred (which can also be done with an analysis of the solutions and the interactions due to $h$). In addition, there is plenty of room for general optimizations, hyperparameter tuning, and different choices (e.g., initial configurations). We have chosen to initialize the nodes in the ground states of $-\hat L^x_i$ since this connects with quantum annealing for qudit systems. However, other initial states, e.g., all colors equally populated, might improve the results further. It will also be important to investigate if the algorithms can provide competitive solutions to GCPs containing millions of nodes and see if they can be generalized to other problems, such as large-scale graph-based machine learning tasks~\cite{zheng2020distdgl,hu2020open}.

Finally, we note that by restricting the cost function to be a quantum Hamiltonian (see, for example, Ref.~\cite{deller_23}), our method is similar to simulating quantum annealing in a qudit system (without entanglement). Therefore, one could also investigate if allowing for entanglement in the system might help the algorithms find better solutions. One approach would be to use tensor networks with a bond dimension larger than 1 (see, for example, Ref.~\cite{bauer_15}). However, this comes with significant computational overhead and might be unfeasible to scale to industrial-size problems.

\textbf{Data availability:} The data supporting our results can be found at \url{https://zenodo.org/records/11483787}.

\section*{Acknowledgment} \label{sec:ack}

 We acknowledge useful discussions with Márcio M. Taddei and Benjamin Schiffer and the support from the Government of Spain (European Union NextGenerationEU PRTR-C17.I1, Quantum in Spain, Severo Ochoa CEX2019-000910-S and FUNQIP), Fundació Cellex, Fundació Mir-Puig, the EU (NEQST 101080086, Quantera Veriqtas, PASQuanS2.1 101113690), the ERC AdG CERQUTE, the AXA Chair in Quantum Information Science and Generalitat de Catalunya (CERCA program). This  project  has  received  funding from the European Union’s Horizon 2020 research and innovation programme under the Marie Skłodowska-Curie grant Agreement No 847517.
 \\
\appendix
\section{Pseudocode}
\label{sec:ap1}
Here, we present pseudocode for the algorithms QdLQA and QdGD (Algorithm~\ref{alg:alg1} and Algorithm~\ref{alg:alg2}, respectively). In both algorithms, we first load the graph's nodes and edges and initialize the hyperparameters and an empty list $e_{\textrm{list}}$.  For each of the $N_{\textrm{runs}}$ of the algorithms, we initialize the state $\psi$ which depends on the parameters $\vec{\phi}$. In practice, this can, for example, be a list of length $\abs{V}$ (the number of nodes), where each entry is a vector and depends on elements of $\vec{\phi}$ as written in Eq.~\eqref{eq:eq1}. In this work, the vectors are initialized to the ground state of $-\hat{L}^x_i$ for QdLQA and random values for QdLQA. However, other initialization schemes are possible. For QdLQA, we then let the time $t$ go from $0$ to $1$, and at each time step, we minimize a cost function $E_{\textrm{Total}}(\psi,t)$ using $\alpha$ gradient descent steps. We then compute the classical energy $e_{\textrm{Potts}}$ (we use-upper case letters for functions and lower-case letters for variables). At the end of each run, the lowest energy is added to $e_{\textrm{list}}$, and finally, $e_{\textrm{list}}$ is returned. This can then be used to extract the values in Table~\ref{table:t1} and Table~\ref{table:t2}. For QdGD, we proceed in a similar manner. Here, however, the cost function does not depend on time, and we also stop the algorithm if no improvement in classical energy is found based on the hyperparameter Patience.

\begin{algorithm}[t]
\label{alg:alg1}
\SetInd{0.2em}{1.3em}
\DontPrintSemicolon
\SetAlgoLined
\caption{QdLQA \\ For the specifics regarding the initialization of the state $\psi$ and formulation of the cost function $E_{\textrm{Total}}(\psi,t)$, see the main text.}\label{alg:cap}
Load $\alpha$, $f$, $N_\text{runs}$, $N_\text{steps}$\;
$e_{\textrm{list}}$ is an empty list \;
\For{$i=1$ \KwTo $N_\text{runs}$}{
    Initialize the state $\psi$ that depend on parameters $\vec{\phi}$ with perturbation $f$\; 
    $t \gets 0$\;
    $e_\text{best}\gets 1000000$\;
    \While{t < 1}{
        Minimize $E_{\textrm{Total}}(\psi,t)$ with $\alpha$ steps\;
        $e_\text{Potts}\gets E_\text{Potts}(\psi)$\;
        $e_\text{best} \gets \text{min}(e_\text{best},e_\text{Potts})$\;
        \If{$e_\text{best}$ = 0} {
            break\;
        }
        $t\gets t + 1/N_\text{steps}$\;
    }
 Add $e_\text{best}$ to $e_{\textrm{list}}$
}
Return $e_{\textrm{list}}$\;
\end{algorithm}

\begin{algorithm}[t]
\label{alg:alg2}
\SetInd{0.2em}{1.3em}
\DontPrintSemicolon
\SetAlgoLined
\caption{QdGD \\ For the specifics regarding the initialization of the state $\psi$ and formulation of the cost function $E_{\textrm{Total}}(\psi)$, see the main text.}\label{alg:cap}
Initialize  $\tilde{f}$, $N_\text{runs}$, $N_\text{steps}$, Patience\;
$e_{\textrm{list}}$ is an empty list \;
\For{$i=1$ \KwTo $N_\text{runs}$}{
    Initialize the state $\psi$ that depend on parameters $\vec{\phi}$  with perturbation $\tilde{f}$\; 
    $e_\text{best}\gets 1000000$\;
    \For{$j=1$ \KwTo $N_\text{steps}$}{
        Minimize $E_{\textrm{Total}}(\psi)$ with one step\;
        $e_\text{Potts}\gets E_\text{Potts}(\psi)$\;
        $e_\text{best} \gets \text{min}(e_\text{best},e_\text{Potts})$\;
        \If{$e_\text{best}$ = 0 \textup{\textbf{or} no improvement }} {
            break\;
        }
        
    }
    Add $e_\text{best}$ to $e_{\textrm{list}}$

}
Return $e_{\textrm{list}}$\;
\end{algorithm}
The pseudocode for the cost function we use for QdLQA is shown in Algorithm~\ref{alg:alg3} (for QdGD it is just the same function evaluated at $t=1$). $\hat L^x$ is a matrix of size $c\times c$ and is given by the $x$ angular momentum operator (we drop the subindex $i$ since it is site independent in our implementation), see Sec.~\ref{subsec:ansatz}, and $j_{\textrm{max}}$ is a fixed qudit initialized to one color. Note that the $j_{\textrm{max}}$ qudit can also be emitted in the second for loop as it only contributes with a constant term. In the algorithm, squaring and taking the logarithm are done componentwise.
Lastly, we show how we extract the final color configuration in Algorithm~\ref{alg:alg4}. There, $conf$ is the zero vector of size $\abs{V}$ and argmax gives the index of the component with the largest magnitude of the vector $\psi[i]^2$.

\begin{algorithm}[t]
\label{alg:alg3}
\SetInd{0.2em}{1.3em}
\DontPrintSemicolon
\SetAlgoLined
\caption{$E_{\textrm{Total}}(t,\psi)$ \\ Computing the cost function from Eq.~\eqref{eq:timedep}. $t$ is a floating point and $\psi$ is, for example, a list of vectors and depends on the parameters $\vec{\phi}$ so that the $i$-th entry is given by Eq.~\eqref{eq:eq1}. }\label{alg:costfunc}
Load vertices $V$, edges $\mathcal{E}$, matrix $\hat L^x$, $j_{\textrm{max}}$, $\gamma$, $h$\;
$e_{\textrm{F}} \gets 0$\;
$e_{\textrm{W}} \gets 0$\;
$e_{\textrm{I}} \gets 0$\;
\For{$(i,j)\in \mathcal{E}$}{
    $h_{i,j}\gets$ sampled from uniform distribution $[0,h)$\;
    $e_{\textrm{F}} \gets e_{\textrm{F}}+ \psi[i]^2 \cdot \psi[j]^2 (\mathcal{E}[i,j] +h_{i,j}) $\; 
}
\For{$l=1$ \KwTo $\abs{V}$}{
    $e_{\textrm{W}} \gets e_{\textrm{W}}+\gamma \psi[l]^2 \cdot \log(\psi[l]^2) $\; 
}
\For{$n=1$ \KwTo $\abs{V}$}{
        \If{$n\neq j_{\textrm{max}}$ } {
        $e_{\textrm{I}} \gets e_{\textrm{I}}-\psi[n]^T  \cdot L^x \cdot \psi[n] $\;
    }
     
}

Return $(1-t)e_{\textrm{I}}+ t(e_{\textrm{F}}+e_{\textrm{W}})$\;
\end{algorithm}

\begin{algorithm}[t]
\label{alg:alg4}
\SetInd{0.2em}{1.3em}
\DontPrintSemicolon
\SetAlgoLined
\caption{$E_{\textrm{Potts}}(\psi)$ \\ Computing the energy from Eq.~\eqref{eq:pottsmodel}. $\psi$ is, for example, a list of vectors and depends on the parameters $\vec{\phi}$ so that the $i$-th entry is given by Eq.~\eqref{eq:eq1}. }\label{alg:costfunc}
Load vertices $V$, edges $\mathcal{E}$\;
$ conf \gets \vec{0}$ \;
\For{$i=1$ \KwTo $\abs{V}$}{
    $conf[i] \gets \textrm{argmax}(\psi[i]^2) $\; 
}
$e_{\textrm{Potts}} \gets 0$\;
\For{$(i,j)\in \mathcal{E}$}{
        \If{$conf[i]=conf[j]$ } {
        $e_{\textrm{Potts}} \gets e_{\textrm{Potts}}+1$\;
    }
}

Return $e_{\textrm{Potts}}$\;
\end{algorithm}

\section{Computer architectures and runtimes}
\label{sec:ap2}
\begin{figure}[t]
\includegraphics[width=0.99\columnwidth]{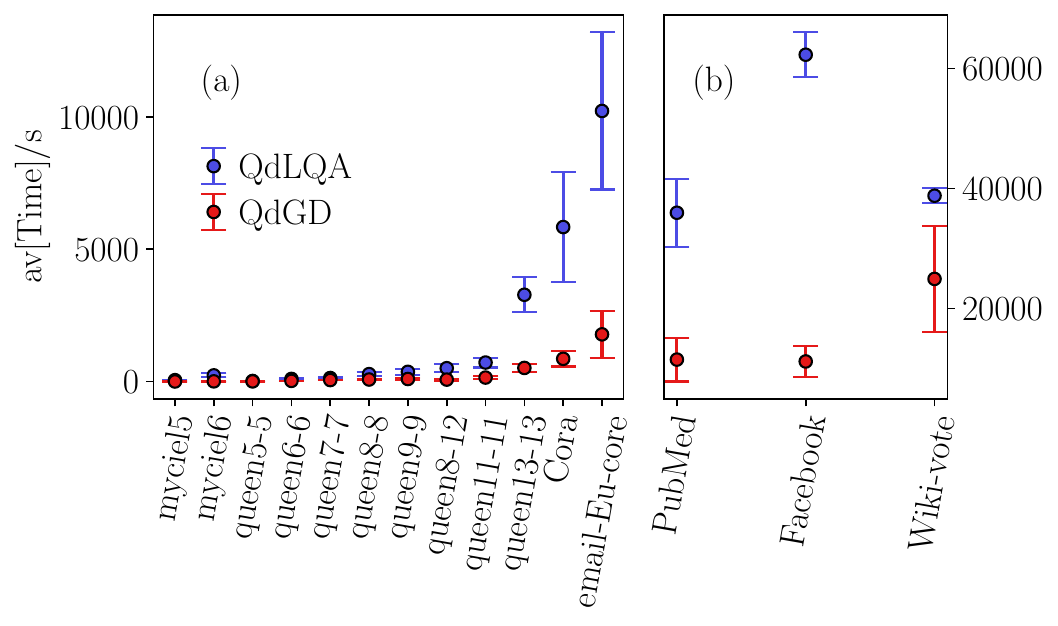}
\caption{Average computation times on (a) a laptop and (b) a stationary computer for $N_{\textrm{runs}}$ of QdLQA and QdGD for different graphs. The error bars correspond to the standard deviation of the data. } 
\label{fig:times}
\end{figure}

All computations, except those on the PubMed, Facebook, and Wiki-vote graphs, were run on a laptop with an 11th-generation Intel(R) Core(TM) i5-1135G7 @ 2.40GHz processor, whereas the latter ran on a desktop with an 11th-generation Intel(R) Core(TM) i5-11500 @ 2.70GHz processor.
In Fig.~\ref{fig:times}, we plot the average computation time for one run for the different graphs for the parameters used in this work; see Appendix~\ref{sec:ap3}. Note that the parameter $\alpha$ varies (for queen13-13), so the figure is an illustration of the cost of obtaining the data presented, rather than fairly comparing the computational cost for the different graphs. Figure~\ref{fig:times}(a) shows the average time for one run of the algorithms for the different graphs on the laptop and Figure~\ref{fig:times}(b) on a stationary computer \footnote{The partially large variances for QdLQA (and possibly for QdGD) comes from the fact that other processes were run simultaneously to the algorithms on the corresponding hardware.}. As expected, the run time is significantly impacted by the Patience hyperparameter for QdGD, which allows the algorithm to stop early.

\begin{figure}[t]
\includegraphics[width=0.99\columnwidth]{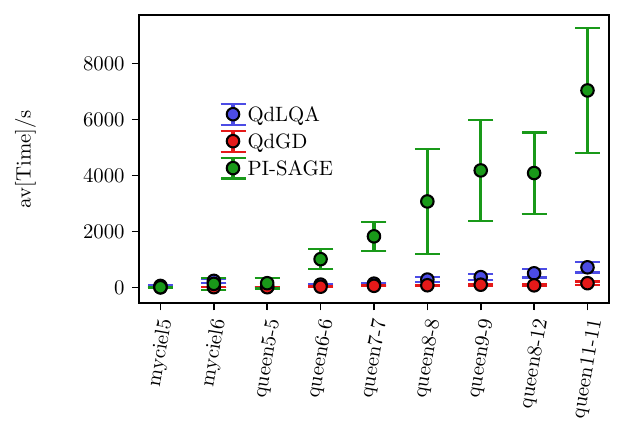}
\caption{Average computation times on the laptop described in the main text for QdLQA, QdGD, and PI-SAGE. For all algorithms, we use $N_{\textrm{runs}}=100$.} 
\label{fig:times_2}
\end{figure}

For a more thorough comparison between the graph neural network approaches, QdLQA, and QdGD, we run the PI-SAGE algorithm from Ref.~\cite{schuetz_22_2} on the previously mentioned laptop for several datasets. For the calculations, we use the code available at~\cite{gnn_git} and the parameters reported in Ref.~\cite{schuetz_22_2}.
In Fig.~\ref{fig:times_2}, we show the average runtimes; as the problem sizes increase, the QdLQA and QdGD run significantly faster (note that this might change on different computer architectures, where, for example, GPUs are available). In addition, the QdLQA and QdGD only optimize $(\abs{V}-1)(c-1)$ parameters (the minus one in $(\abs{V}-1)$ comes from already assigning one color to one qudit in the initial state), which is a fraction of what, for example, PI-SAGE requires (even though the number of parameters in principle scales as $\sim c+\abs{V}$). For example, for the queen5-5, queen9-9, and queen11-11 datasets, our methods have 96, 720, and 1200 parameters, whereas the PI-SAGE method uses 7210, 12663, and 13411 parameters (using the hyperparameters from Ref.~\cite{schuetz_22_2}).

Lastly, we emphasize that the graph neural network approaches also might need $N_{\textrm{runs}}>1$ to find an optimal solution. By introducing $p^{\textrm{PI-SAGE}}_{\textrm{min}}=N_{\textrm{min}}/N_{\textrm{runs}}$, we get the following tuples (dataset, $E^{\textrm{PI-SAGE}}_{\textrm{min}}$, $p^{\textrm{PI-SAGE}}_{\textrm{min}}$): (myciel5, 0, 100/100), (myciel6, 0, 100/100), (queen5-5, 0, 85/100), (queen6-6, 0, 17/100), (queen7-7, 0, 1/100), (queen8-8, 1, 3/100), (queen9-9, 1, 6/100), (queen8-12, 2, 6/100), and (queen11-11, 13, 1/100) in our calculations (the last result actually improves on what was reported in Ref.~\cite{schuetz_22_2}).

\section{Hyperparamters}
\label{sec:ap3}
\begin{center}
\begin{table*}
\renewcommand{\arraystretch}{1.4}
\begin{tabular}{ |c|c|c|} 
 \hline
 Hyperparameter &Algorithm & Description:  \\ 
 \hline
  $N_\text{steps}$ &   QdLQA &Number of steps used when going from $t=0$ to $t=1$ in the quantum adiabatic evolution  \\ 
 \hline
   $N_\text{steps}$ &   QdGD &Number of gradient descent steps  \\ 
 \hline
 $\gamma$ & Both &Prefactor in $E_{\textrm{W}}(\psi)$ from Eq.~\eqref{eq:weight} \\ 
  \hline
  $\alpha$ & QdLQA &Determines the number of gradient descent steps at each time step $t$ \\
   \hline
    $\eta$ & Both &Learning rate in the Adam optimizer \\ 
     \hline
      $f$ &QdLQA & Determines the perturbation in the angles in the initial state \\ 
       \hline
            $h$ &Both & Determines the perturbation in the interactions in the Potts model, see Eq.~\eqref{eq:final} \\
             \hline
                  $N_\text{runs}$ &Both & Number of times we run the algorithm \\ 
                     \hline
                        Patience &QdGD & If no improvement in $E_{\textrm{Potts}}$ is seen after this many gradient descent steps, the algorithm stops \\ 
                           \hline
                               $\Tilde{f}$ &QdGD & Determines the sampling interval for each qudit in the initial state \\ 
                 
 \hline
\end{tabular}
\caption{Summary of the hyperparameters for the QdLQA and QdGD algorithms. }
\label{table:t4}
\end{table*}
\end{center}

\begin{center}
\begin{table*}
\renewcommand{\arraystretch}{1.4}
\begin{tabular}{ |c|c|c|c|c|c|c|c|c|c| } 
 \hline
  & \multicolumn{7}{c|}{QdLQA }   \\ 
 \hline
 Graph & $N_\text{steps}$ & $\gamma$& $\alpha$ & $\eta$ & $f$  &$h$   & $N_\text{runs}$  \\ 
 \hline
  myciel5 & 1000 &1.0 &1& 0.5 & 0.0 & 3.0 & 100 \\ 
 \hline
   myciel6 & 1000 & 1.0 & 1& 0.5 & 0.0 & 3.0  & 100\\ 
 \hline
 queen5-5 & 1000 & 1.0 & 1& 0.5 & 0.0 & 3.0 & 100 \\ 
 \hline
 queen6-6 & 1000 & 1.0 & 1& 0.5 & 0.0 & 3.0  &100  \\ 
 \hline
 queen7-7 & 1000 & 1.0 & 1& 0.5 & 0.0 & 3.0  &100  \\ 
 \hline
 queen8-8 & 1000 & 1.0 & 1& 0.5 & 0.1 & 3.0  &100  \\ 
 \hline
 queen9-9 & 1000 & 1.0 & 1& 0.5 & 0.0 & 3.0  &100 \\ 
 \hline
 queen8-12 & 1000 & 1.0 & 1& 0.5 & 0.0 & 3.0  &100 \\ 
 \hline
 queen11-11 &1000 & 1.0 & 1& 0.5 & 0.1 & 3.0  &100   \\ 
 \hline
 queen13-13 & 1000 & 0.75 & $*$& 0.5 & 0.0 & 3.0  &100  \\ 
 \hline
   Cora &  1000 & 0.25 & 1& 0.5 &  0.1 &10.0 &100  \\ 
 \hline
   Pubmed & 1000 &0.1 & 1& 0.1 &  0.5 &3.0 &100 \\ 
   \hline
      email-Eu-core & 1500 &0.5 & 1& 0.1 &  0.1 &3.0 &100 \\ 
   \hline
 Facebook & 1000 &0.1 & 1& 0.1 &  0.1 &3.0 &100 \\ 
 \hline
  Wiki-vote & 1000 &0.1 & 1& 0.5 &  0.1 &3.0 &100 \\ 
  
 \hline
\end{tabular}
\begin{tabular}{ |c|c|c|c|c|c|c|c|c|c| } 
 \hline
 & \multicolumn{7}{c|}{QdGD }   \\ 
 \hline
 \, \, \,& $N_\text{steps}$ & $\gamma$& Patience & $\eta$ & $\tilde{f}$  &$h$   & $N_\text{runs}$  \\ 
 \hline
   & 1000 &1.0 & 100& 0.5 & 1.0 & 3.0 & 100 \\ 
 \hline
    & 1000 & 1.0 &100& 0.5 & 1.0 & 3.0  & 100\\ 
 \hline
  & 1000 & 1.0 &100& 0.5 & 1.0 & 3.0 & 100 \\ 
 \hline
  & 1000 & 1.0 & 100& 0.5 & 1.0 & 3.0  &100  \\ 
 \hline
  & 1000 & 1.0 & 100& 1.0 & 1.0 & 1.0  &100  \\ 
 \hline
  & 1000 & 0.5 & 100& 1.0 & 1.0 & 1.0  &100  \\ 
 \hline
  & 1000 & 0.5 & 100& 0.1 & 1.0 & 1.0  &100 \\ 
 \hline
  & 1000 & 1.0 & 100& 0.5 & 1.0 & 3.0  &100 \\ 
 \hline
  &1000 & 1.0 & 100& 0.5 &1.0 & 3.0  &100   \\ 
 \hline
  & 3175 & 1.0 & 100& 0.1 & 1.0 & 3.0  &100  \\ 
 \hline
    &  1000 & 0.5 & 100& 1.0 &  1.0 &3.0 &100  \\ 
 \hline
      & 1000 &0.5 & 100& 0.5 &  1.0 &3.0 &100 \\ 
   \hline
    & 1000 &0.5 & 100& 0.1 &  1.0 &1.0 &100 \\ 
   \hline
  & 1000 &0.1 & 100& 0.1 &  1.0 &3.0 &100 \\ 
 \hline
   & 1000 &0.1 & 100& 0.5 &  1.0 &3.0 &100 \\ 
  
 \hline
\end{tabular}
\caption{Hyperparameters used to compute the data in Tables~\ref{table:t1} and~\ref{table:t2}. $*$ For queen13-13, we conduct $\textrm{round}(e^{2 t })$ gradient  descent steps ($\textrm{round}()$ indicates that we round to the nearest integer) at each point in time (maximum 7). }
\label{table:t5}
\end{table*}
\end{center}
There are several hyperparameters needed in both the QdLQA and the QdGD algorithms. These and their effects are summarized in Table~\ref{table:t4}. In addition, the Adam optimizer has further hyperparameters that could be tuned (here, we only adjust the learning rate), however, in this work, they are left to their default values in PyTorch.

In Table~\ref{table:t5}, we show the hyperparameters used to produce the data in Tables~\ref{table:t1} and~\ref{table:t2}. Initially, we choose the default values $\eta=0.5$, $h=3.0$, $\alpha=1$, $N_{\textrm{steps}}=1000$ and $\gamma=1.0$ which work well for many of the graphs. When the algorithms neither solved the problem nor achieved better solutions than the graph neural network approaches, we either tuned the parameters manually or did a grid search. Note that for queen13-13, we choose $N_{\textrm{steps}}=3175$ for QdGD so that we get equally as many gradient descent steps as in QdLQA.

 \bibliographystyle{biblev1}
 \bibliography{references}

\begin{thebibliography}{10}
\bibitem{Note*}
\bibinfo{note}{The partially large variances for QdLQA (and possibly for QdGD) comes from the fact that other processes were run simultaneously to the algorithms on the corresponding hardware.}
\end{thebibliography}
\end{document}